\documentclass[12pt,a4paper]{article}
\usepackage{epsfig}
\input epsf
\epsfverbosetrue
\begin{document}
\title {\begin{flushleft}{\large \tt LC-DET-2006-008\\ November 2006}\end{flushleft}
\vspace{1cm}
\rm \bf\LARGE  A proposed DAQ system for a calorimeter at the International 
Linear Collider\\
\vspace{1cm}}

\date{}                    
\maketitle

\begin{center}{M.~Wing$^{1,\dagger}$, M.~Warren$^1$, P.D.~Dauncey$^2$ and J.M.~Butterworth$^1$ \\for CALICE-UK groups}

\bigskip
{\small $^1$University College London}, {\small $^2$Imperial College London}

{\small $^\dagger$Contact: mw@hep.ucl.ac.uk}
\end{center}

\begin{abstract}
\noindent

This note describes R\&D to be carried out on the data acquisition system for a 
calorimeter at the future International Linear Collider. A generic calorimeter 
and data acquisition system is described. Within this framework modified designs 
and potential bottlenecks within the current system are described. Solutions 
leading up to a technical design report will to be carried out within 
CALICE-UK groups.

\end{abstract}

\pagestyle{plain}
\thispagestyle{empty}
\clearpage

\pagenumbering{Roman}                                                           

\newpage
\setcounter{page}{1}
\setcounter{figure}{0}
\setcounter{section}{0}
\setcounter{subsection}{0}
\pagenumbering{arabic}
%


\section{Introduction}

With the decision on the accelerator technology to be used for a future 
International Linear Collider (ILC), detector R\&D can become more focused. The 
time-line for an R\&D programme is also clearer with, assuming a technical design 
report to be written by 2009, three years to define the make-up of a given 
sub-detector. Within the CALICE collaboration, which is designing a calorimeter 
for the ILC, a collection of UK groups (CALICE-UK) are part of the initial effort 
to prototype a calorimeter composed of silicon and tungsten~\cite{tdr}. The 
electromagnetic section of the calorimeter (ECAL) has been taking test-beam data  
at DESY and CERN in 2006. The UK has designed and built electronics to 
readout the ECAL~\cite{previous_bid} - these are also now being used by the analogue 
hadronic calorimeter - and is taking part in the current  
data-taking period. Building on this expertise, CALICE-UK has defined an R\&D 
programme. A significant part of this programme is the design of the data acquisition 
(DAQ) system for a future calorimeter. 

In the work, DAQ equipment will be developed which attacks likely bottlenecks 
in the future system and is also sufficiently generic to provide the readout for 
new prototype calorimeters, such as the prototype to be built in the EUDET 
project~\cite{eudet}. The main aim is to start an R\&D programme which will 
work towards designing the \emph{actual} DAQ system of the future calorimeter. 
Alternative designs of a DAQ system which could affect the layout of the final 
detector or functionality of components are also considered. 
The concept of moving towards a ``backplaneless'' readout is pursued. A strong 
under-pinning thread here is to attempt to make use of commercial components and 
identify any problems with this approach. Therefore the system should be easily 
upgradable, both in terms of ease of acquiring new components and competitive prices. 

This note is organised as follows. The parameters of the superconducting 
accelerator design and calorimeter structure and properties which impinge upon 
considerations of the DAQ system for a calorimeter are discussed in 
Section~\ref{sec:param}. The main body of the note in Section~\ref{sec:design} 
discusses the DAQ design and proposes areas of R\&D within it. The work is 
will investigate the three principal stages of the DAQ 
system: the connection along the calorimeter module; the connection from the on- to 
off-detector; and the off-detector receiver. In Section~\ref{sec:model} a model 
DAQ system for the final ECAL is proposed. This necessarily makes many assumptions 
but gives an idea of the scale of the system involved: it can also be the start 
of an initial costing. The note ends with a brief summary in 
Section~\ref{sec:summary}.

The programme detailed below will allow CALICE-UK groups to continue to assist
in development of new technologies for the DAQ system. We would expect to write a 
chapter in the future technical design report on the DAQ system for the 
calorimeter. For the final calorimeter, the DAQ should ideally be the same for the 
ECAL and HCAL.  Although CALICE-UK has so far concentrated on the ECAL, our 
proposals for R\&D contained in this document are sufficiently generic that both 
calorimeter sections should be able to converge to use the DAQ system we design. 
This will place us in a position to build the DAQ system for future 
large-scale prototype calorimeters (e.g. EUDET) and the final system. Indeed the 
principle of a generic design using commercial components should be applicable to 
many detector sub-systems. Therefore, the R\&D to be performed here may have 
consequences or applications to the global DAQ system for a future detector.

\section{General detector and accelerator parameters}
\label{sec:param}

\begin{figure}[htp]
\begin{center}
~\epsfig{file=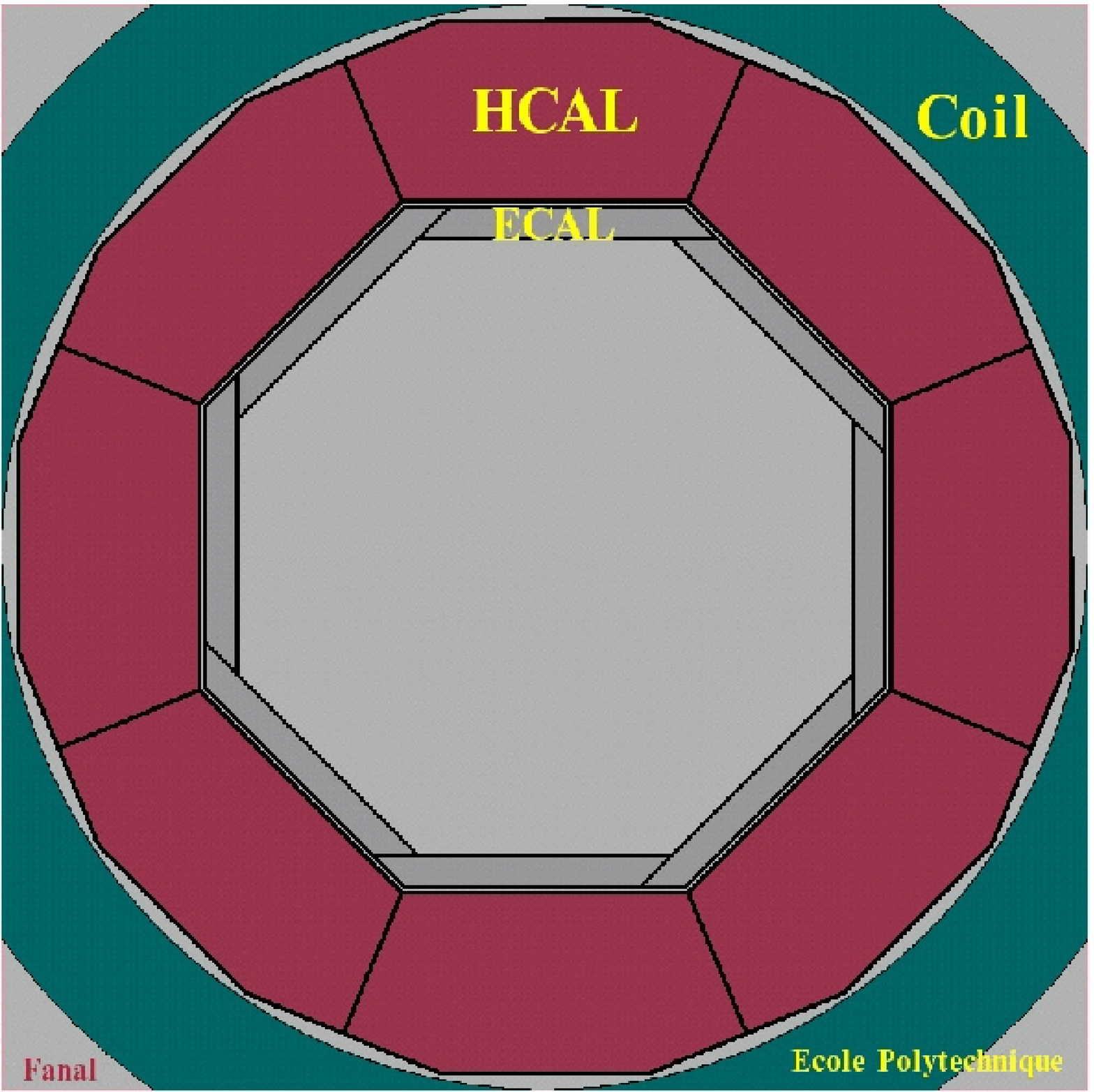,height=7.5cm}
~\epsfig{file=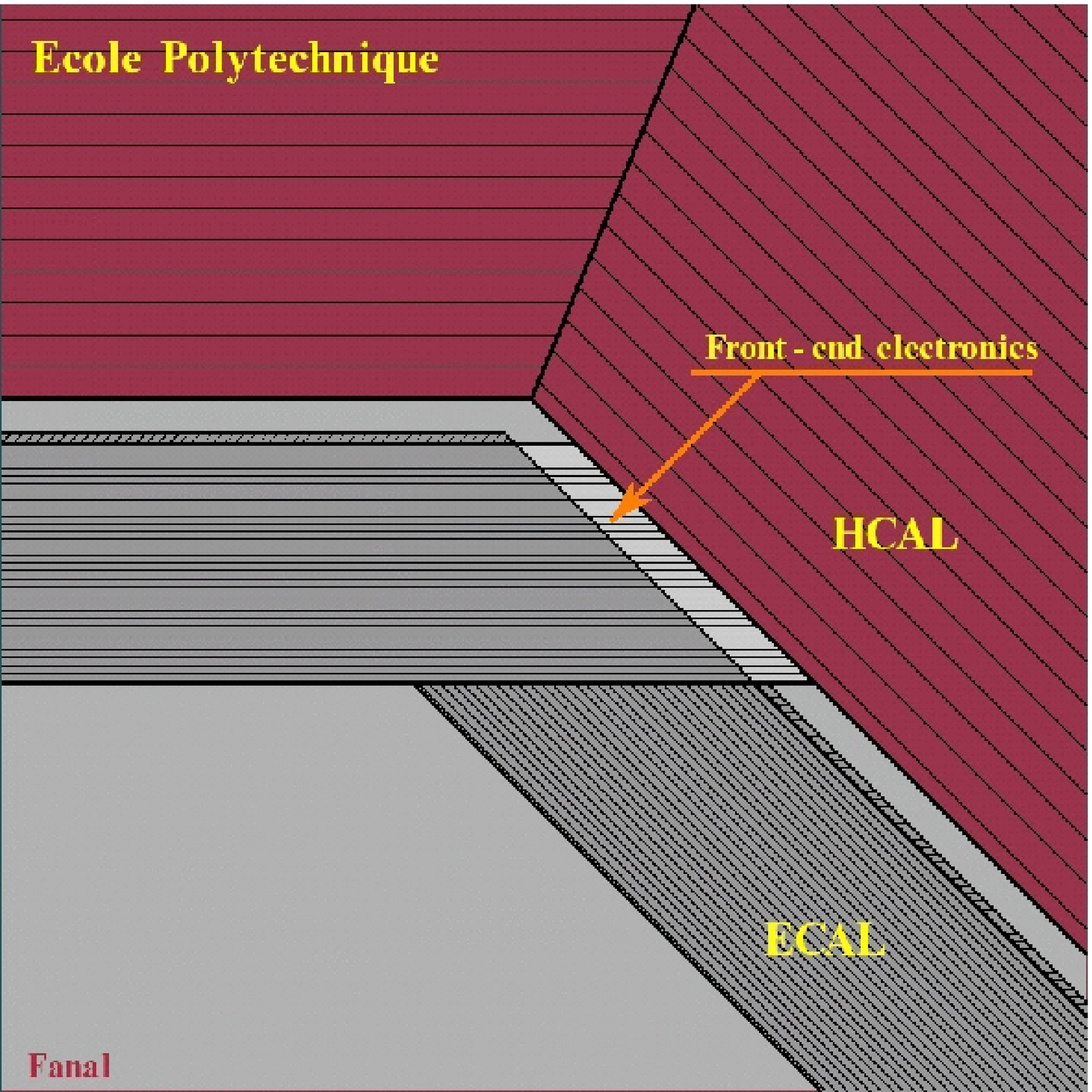,height=7.5cm}
\caption{View of the barrel calorimeter modules and detail of the overlap region 
between two modules, with space for the front-end electronics.}
\label{fig:cal}
\end{center}
\end{figure}

The design~\cite{tdr} for a calorimeter for the ILC poses challenges to the 
DAQ system mainly due to the large number of channels to be 
read out. The TESLA design~\cite{tdr} for a sampling electromagnetic calorimeter is 
composed of 40 layers of silicon interleaved with tungsten. The calorimeter, shown 
in Fig.~\ref{fig:cal}, has eight-fold symmetry and dimensions: a radius of about 
2\,m, a length of about 5\,m and a thickness of about 20\,cm. Mechanically, the 
calorimeter will consist of 6000 slabs, of length 1.5\,m, each containing about 
4000 silicon p-n diode pads of 1$\times$1\,cm$^2$, giving a total of 24~million 
pads. More recent designs for the detector collaborations consider fewer layers, 
29 for LDC~\cite{ldc-outline} and 30 for SiD~\cite{sid-outline}, and also smaller pad sizes of 
5$\times$5\,mm$^2$, or even 3$\times$3\,mm$^2$.

A generic scenario for the DAQ system is as follows. At the very front end (VFE), 
ASIC chips will be mounted on the PCBs and will process a given number of pads. The 
ASICs will perform pre-amplification and shaping and should also digitise the data 
and {\em may} even apply a threshold suppression. The current 
design~\cite{de_la_taille:priv:comm} of such chips has each containing 64~channels, 
although this may be higher in the final calorimeter. The ASIC power consumption 
has to be minimised as they are difficult to cool due to the small gaps between 
layers which are required to take advantage of tungsten's Moli\`{e}re radius. The 
data will be digitised in the ASIC and transferred to the front-end (FE) 
electronics which are placed in the detector at the end of the slab as shown in 
Fig.~\ref{fig:cal}. It is expected that zero suppression will be done in the FE 
(using FPGAs) to significantly reduce the rate. The data will then be transferred 
off the detector, probably via a network switch, to a receiver of many PCI cards in 
a PC farm. 

If we assume the TESLA design for data taking at 800~GeV, the 
following parameters have to be considered. There will be a 4886 bunch crossings 
every 176~ns in a bunch train, giving a bunch train length of about 860~$\mu$s. The 
bunch train period is 250~ms, giving a duty factor between trains of about 0.35\%. 
The ECAL is expected to digitise the signal every bunch crossing and readout 
completely before the next bunch train. In a shower, up to 100~particles/mm$^2$ can 
be expected, which in a 1$\times$1~cm$^2$ pad equates to 10000 minimum ionising 
particle deposits. The ADC therefore needs a dynamic range of 14~bits. Assuming no 
threshold suppression and that 2 bytes are used per pad per sample, then the raw data 
per bunch train is $24 \cdot 10^6~\times$~4886~$\times$~2~=~250~GBytes which equates to 
0.3-2.5~MBytes for each ASIC depending on whether they process between 32 and 256 
channels. The data appears within a bunch train length of 860~$\mu$s giving a rate out 
of the ASIC of 0.4-3~GBytes/s, which we take to be 1~GBytes/s from now on. Threshold 
suppression and/or buffering (to allow readout between bunch trains) within the 
ASIC could reduce this rate. However, suppression in the ASIC may not be flexible 
enough compared with doing this in the FE and buffering requires some ASIC power to 
remain on between bunch trains, potentially generating too much heat. Hence the 
rates after the VFE depend on the assumptions made and system layout and will be 
discussed for each individual case where necessary.

\section{Design of a DAQ system}
\label{sec:design}

\subsection{Transmitting digitised data from the VFE chip}

The transmission of digitised data from the ASIC is very heavily influenced by what can 
be done within the slab given the low heat-load requirements due to the difficulties of 
cooling. It is not yet known what the capabilities of the VFE ASIC will be, so various 
possibilities were considered.

In general, somewhere in the readout system, there will have to 
be an ADC and a threshold discriminator. These tasks could in principle be performed 
in either order and could be done in the VFE or in the FE. There is also the 
possibility of buffering events in either the VFE or FE. This would allow the data to 
be read out between bunch trains 
rather than bunch crossings. This entails a dramatic decrease in the rate 
of read out due to the large spacing between bunch trains. There is then a matrix of 
possibilities, with some number 
of the functionalities, ADC, thresholding and buffering, being done in the VFE or FE. 
Below the four possibilities are considered for the ADC and thresholding, and 
also the buffering in the VFE.

\begin{enumerate}

\item Neither ADC nor thresholding is done in the VFE

\item Only the ADC is done in the VFE

\item Only the thresholding is done in the VFE

\item Both are done in the VFE

\item Buffering done in the VFE

\end{enumerate}

We consider that threshold discrimination is best done after 
the ADC step rather than before. This allows much easier monitoring the 
pedestals and noise, etc., by allowing some readout at a low rate even when below 
the threshold. In addition, setting a stable analogue threshold is not easy; any 
drifts will change the level. The uniformity over all channels might not be good 
enough which would then require a large number of trim DACs. 

1) If neither an ADC or threshold discriminator is built into the VFE ASIC (due to 
them taking too much power), then the raw analogue signals will be sent out to the 
FE. This is 2k analogue channels which require around 14 bits precision, which is 
not trivial to achieve. Even if this can be done, digitising the data at the 
FE would be hard. The space is limited and so it is likely only a restricted number 
of ADCs could be mounted in this area. Assuming 20 channels of ADCs would be 
possible, then each would have to handle 100 pads, with these being multiplexed in 
turn into the ADC. To keep up with the sampling rate needed, i.e. 176~ns for each 
channel would therefore require the ADCs to sample at 1.76~ns. Finding a 14-bit FADC 
which can do this would not be easy. The 
alternative would be to use an analogue pipeline; assuming one for each of the 
20~ADC channels would result in each pipeline storing about 500k analogue 
samples which is difficult. Putting an analogue threshold in front of the ADCs would 
clearly cut the rate down but would need a major ASIC development to handle this; a 
variable length analogue pipeline with time-stamps would be needed. This is in 
addition to the pedestal monitoring problems mentioned above.

2) Only doing the ADC on the VFE seems a much more reasonable option. The 14-bit 
requirement is much easier to achieve with a short signal path before the ADC. The 
digitised data can be transmitted from the VFE to the FE more easily than analogue 
data. The rates are not trivial however; these would be around 50~GByte/s per slab, 
or 1~GByte/s from each wafer/ASIC. This is at the level where a fibre would be needed; 
commercial fibres now carry 5~GBytes/s. Fibres are also less noisy than copper. This 
use of fibres within the slab would raise many other issues such as the power needed 
to transmit the light out (or could it be supplied by an external laser and then only 
modulated on the ASIC), how to reliably attach the fibres at each end (a total of 
300~000 fibres would be needed for 6000 slabs each with 50 ASICs), how large the fibre 
connectors would be (the total thickness within the slabs is limited to some mm only), 
etc.. Although this is an active area of commercial development, it is not clear if 
opto-electronic intra-PCB communications will become standard enough on the 
time-scale needed~\cite{chip_to_chip}.

It is clear some development would be needed for this to be an option; the equivalent 
system in ATLAS has three fibres transmitting a total of 10~MBytes/s with a 2~mm high 
connector needed. Self-aligning silicon-fibre interfaces are possibilities; while we 
could not do significant R\&D compared with the commercial sector, we could test 
industrial prototypes and do R\&D in conjunction with industry.

Once the data are on a fibre direct from the ASIC, the idea of whether any FE 
electronics would be needed at all was raised, as the fibre would go 10s of metres, 
bypassing the FE completely. However, shipping out all the raw data to the offline 
seems an expensive overkill, but is considered as this may change with commercial 
development. 

3) Only doing the threshold in the VFE suffers from the same problems as mentioned above; 
there is a difficulty of 
monitoring the pedestals as well as the complexity of the ASIC needed to handle the 
channels. 

4) Doing both ADC and threshold in the VFE places the easiest requirements on the FE, 
with a corresponding increase in difficulty for the VFE. Assuming the threshold is 
applied after the ADC, some communication of the threshold and other configuration data 
from the FE to the VFE will still be needed. The data rate out is clearly reduced; it 
would be around 400~MBytes/s for the slab, or 20~MBytes/s for each wafer/ASIC. Although 
easiest for transferring data from the VFE to FE, due to the low rates, it is not clear 
if the threshold can be reliably applied in the VFE. This scenario also looks like the 
situation if the Monolithic Active Pixel Sensors (MAPS) technology - essentially a 
digital calorimeter - were used rather than silicon diodes. For the diode 
option in this scenario, it is also questionable as to whether significant FE electronics 
logic is needed. As the ASIC performs 
both ADC and threshold suppression, the data could be transferred directly off the 
detector.

5) It is generally assumed that buffering in the FE is possible, with large  
amounts of memory available in modern FPGAs. However, the issue of buffering in 
the ASICs is more technically challenging. The challenges for such a procedure 
are having a large enough memory integrated into the ASICs and the keeping the 
power low whilst the data is being read out between bunch trains. The advantages 
are clear: the rate of transmission from the ASIC to the FPGA is reduced by about 
two orders of magnitude. For the proposed electrical connections along the board 
this will ease the transmission significantly.

\begin{figure}[htp]
\begin{center}
~\epsfig{file=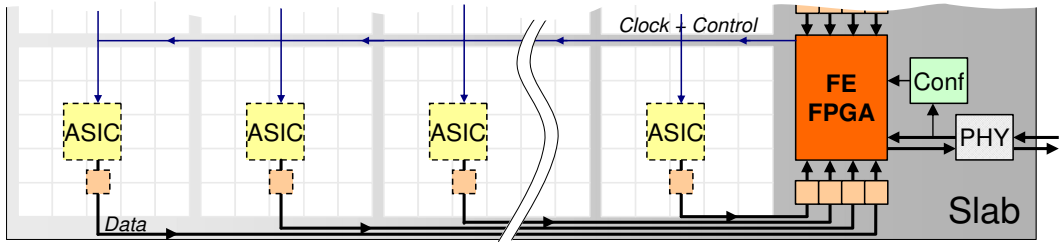,height=4.5cm}
\caption{Design of VFE to FE link.}
\label{fig:vfe-to-fe}
\end{center}
\end{figure}

We, therefore, propose R\&D for two scenarios where only the ADC is done in the VFE and 
where the ADC and thresholding are done in the VFE, both coupled with buffering in the 
VFE because they 
provide realistic solutions and have complementary applications. In favour of only 
performing the ADC,  any 
threshold suppression can be performed more accurately in the FPGA at the FE rather 
than in the VFE. When thresholding is also done in the VFE along with the ADC, the 
data transfer rate from the VFE and FE is significantly 
smaller. A schematic of scenario 2) is shown in Fig.~\ref{fig:vfe-to-fe}.

In both scenarios, we intend to set-up a mock data transfer system which requires 
having a test board with FPGAs linked by fibres. This will simulate a link between 
the VFE ASICs and the FE FPGAs. Any developed system e.g. a new VFE chip design or 
the MAPS set-up could also be tested in our prototype system. We will also 
demonstrate that the system would work for the hadronic calorimeter as well as 
the ECAL. This would require modifying the system to have a more links but a lower 
rate. The prototype will 
incorporate, wherever possible, commercially available components such as 
Virtex-4~FPGAs~\cite{virtex4-fpga} which has multi-gigabit serial transceivers and 
is compatible with 10/100/1000 Mbit/s ethernet and PCI express x16 and higher.


The final chip should have around 64 channels and would be embedded inside the detector. 
The ADC(s) should be included in the chip in order to output digital data serially 
at high rate (typically 1-2Gbit/s). The DAQ would thus look more like ``an event 
builder'' than a traditional DAQ. It would perform the data reformatting (from 
``floating'' gain + 10bit to 16 bit), calibration, possibly linearisation and some 
digital filtering. It is possible that at this level, some event processing will be 
performed. The other task of the DAQ is to load all the parameters needed by the 
front-end, control the power cycling and run the calibration. These specifications 
fit in well with our current generic system.

The current version of the VFE ASIC chip~\cite{vfeasic} is being used to read out 
the existing CALICE ECAL. This chip does not meet the requirements for the ILC ECAL 
and the development of the design is an ongoing project in 
LAL/Orsay~\cite{de_la_taille:priv:comm}. In the next 1--2~years, it is expected to 
have a version of such a chip with low enough power and noise that would serve as a 
realistic prototype. This ASIC is expected to have (at least) 32~channels, an internal ADC per 
channel, multiple gain ranges, and optional threshold suppression and digital 
buffering to reduce the required output rate. 

Instead of using silicon diodes, the feasibility of using the MAPS technology 
is to be investigated~\cite{maps_lcnote}. The use of this technology would also have 
an impact on the design of the DAQ system. Here, there would be no ADC and a threshold 
has to be applied on the wafer, by definition. The data rate for a final detector 
would be 3~GBytes/s per slab, or 150~MBytes/s per wafer, which is low enough for 
non-fibre communication. 

If threshold suppression or buffering could be done in the VFE ASIC, the rate to 
the FE would be reduced by two orders of magnitude. Current designs cannot do this 
and it may not even be desirable or practical, so we have to allow for data rates 
of order GByte/s needing to be transferred out of each VFE ASIC during the bunch 
train. Whether an electrical or optical connection would be needed has to be 
investigated. Although chip-to-chip fibres are not yet standard technology, this 
is an active area of industrial research~\cite{chip_to_chip}.

Issues of how the data 
would be transported from the VFE to FE have to be considered and can be done 
already without a real prototype. Transporting of order GByte/s of data over 
1.5\,m in a very limited space is a challenge. The conventional approach would be 
to use copper but here the effects of noise and interference will have to be 
considered. There is also the possibility of using optical fibres although here 
there are also design considerations: the size of connectors would have to be 
investigated as the vertical clearance at the VFE is of the order of mm and the 
power needed to transmit light out would also need to be investigated. This work 
ties in closely with the mechanical and thermal aspects of the design. 

In preparation for a real prototype, a test system will be built with a 1.5~metre 
PCB containing FPGAs linked optically or electronically. The data transfer would 
then be considered as a function of the number of VFEs, whether zero suppression is 
done in the VFEs and whether data is buffered during the bunch train. The bandwidth 
and cross-talk of the data transfer can be simulated using CAD tools. The clock 
and control distribution from the front-end to the VFE chips can be investigated 
as to whether one transmission line per chip is needed or multi-drop is possible. 

\subsection{Connection from on- to off-detector}

In this section, we consider two widely differing scenarios, (shown in 
Figure ~\ref{fig:data_topol}): 
\begin{figure}[ht]
\begin{center}
~\epsfig{file=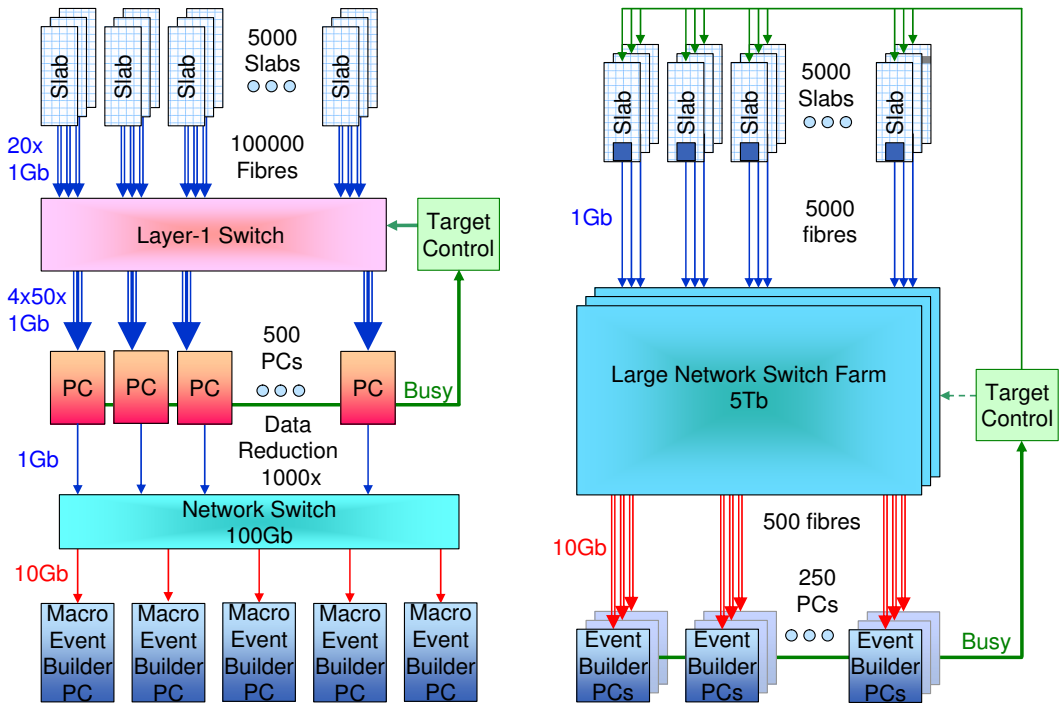,height=11cm}
\caption{A comparison of the two scenarios, ``Alternative configuration'' (left) 
and ``Standard configuration'' (right) for the on- to off-detector DAQ system.}
\label{fig:data_topol}
\end{center}
\end{figure}

{\bf Standard Configuration}\\
In our assumed standard detector configuration, communication from the VFE will pass 
via the electronics at the FE to an off-detector receiver. We assume that threshold 
suppression will be done at the FE, and hence the rate would be significantly 
reduced from that at the VFE. Assuming that the rate is reduced to 1\% of the 
original data volume of 250\,GBytes per bunch train and each sample above threshold 
needs a channel and timing label, the total data volume to be read out from the 
calorimeter is about 5\,GBytes or about 1\,MByte per slab. These data have to be 
read out within a bunch train period of 250\,ms, giving a rate of 5\,MBytes/s.

{\bf Alternative configuration}\\
Here we imagine that the FE is removed and the communication is directly from the 
VFE to the off-detector. We assume that the ASIC only digitises the data and 
250\,GBytes has to be transported off the detector per bunch train. This will 
require a high-speed optical network. It should be noted, however, that the need for 
FE electronics also becomes questionable if more processing is done on the ASIC chip, 
such as threshold suppression. In such a scenario, transporting the data directly 
from the ASIC off the detector could be done and so the FE would become redundant. 
The number of fibres required to read-out the 24~million channels would vary between 
750\,000 to about 90\,000 depending on whether the ASIC handles 32 or 256 channels.
If we assume that the diameter of a fibre is 150\,$\mu$m, or with cladding 
250\,$\mu$m then if half of the circumference of 12\,m had fibres running along it, 
the bundle would be up to 1\,cm in depth, but could be as little as 1\,mm, depending 
on the number of ASICs and hence fibres. This would leave ample room for other cables 
and power supplies. This concept would revolutionise the whole calorimeter design and 
so needs to be considered now when changes in its general structure could be 
considered. Our research in this area will provide important feedback to groups 
designing the ASIC chips.

The off-detector receiver, as described later, is assumed to consist of PCI cards 
housed in PCs. The reliability of large PC farms is an issue for reading out the data; 
if one PC goes down, all of the data in that region of the calorimeter is lost. Current 
PC farms show a rate of 1 PC failure per day in a farm of 200. This is not large but is 
also not small and would require some surplus of PCs (say 10\%) above the number 
required based just on the number of detector channels. For a final working calorimeter 
readout system these PCs would need to be repaired and put back into the system on a 
regular basis. The standard scenario would require less high-speed 
equipment off the detector, whereas the alternative would require many optical fibres 
with dedicated optical switching. The alternative scenario would, however, remove 
material from inside the detector which would ease construction and have a potentially 
advantageous impact on event reconstruction and, hence, physics measurements. It would 
also reduce the number of processing components within the detector which could be 
attractive since they would be inaccessible for long periods of time.

Using the experience gained from the two scenarios above, a hybrid path can be explored.
By optimising the functionality in the VFE and matching it to an FE, overall 
instantaneous data rates can be reduced as well as lowering fibre count. For example, 
if data is buffered in the VFE, some form of passive multiplexer can be envisaged that 
will combine data from all the VFEs. This could take the form of spliced optic-fibres, or 
an OR gate. This would not necessarily remove the need for an FE, but the reduced size 
and complexity will have many benefits (thermal, configuration time, SEUs, cost, etc.). 

To transmit data onto the detector, we will attempt to use the same commercial hardware 
used for off-detector communication. The requirements are different, though, as the  
detector front-end requires clock and synchronisation signals as well as low-level 
configuration and data signals. Commercial network hardware is not ideally suited to 
synchronous tasks but the significant cost and reliability benefits make it worthy of 
investigation. This is split into two areas: failsafe configuration prior to normal 
running, and clock and control signal distribution.

{\bf Failsafe Configuration}
In a scenario requiring an FPGA on-detector, it is imperative that the device can be 
re-configured remotely. This is necessary not only because of the number of FPGAs but 
also because of the uncertainty in the detector performance in terms of data 
suppression, pedestal drifts, bad channel identification etc., all of which have to be 
implemented in the FE FPGA. The optimal algorithms for these tasks will only be 
determined after some experience of operating the calorimeter. As the \mbox{FPGAs} are 
relatively inaccessible, it is important that a failsafe method of reseting and 
re-configuring the devices exists.

Avoiding the use of additional control lines necessitates a means to extract 
signals from the communications interface. Most probably this will involve 
``spying'' on the serial data line.

Of primary importance is to force the FPGA into a known state under any conditions 
(i.e. a hard-reset). It may simply be possible to send an exceptionally long series of 
``1''s to the slab (say 40M = 1 second) where an RC type-circuit with a long time-constant 
will trigger the reset. A more complex method would require some discrete logic (like a 
shift-register with parallel output into a comparator) searching the incoming data-stream 
for a ``magic'' number. Power cycling the board is also a (less elegant) solution, but 
still requires attention in circuit design to ensure the FPGA will boot.

Once hard-reset, the FPGA will follow its start-up logic to initiate the boot - making use 
of 2 distinct methods:

\begin{itemize}

\item Use a non-volatile base configuration that is either hardwired into the FPGA or 
provided by a EEPROM external to the FPGA. In this case the FPGA would assist in writing 
the configuration to its internal RAM and then issue a re-boot-using-internal-configuration 
command to itself. 

\item By waiting for external stimulus (usually a clock) to control data the data being fed 
into the device. This requires additional hardware to format the the serial-data into 
data/clock/control for FPGA consumption.

\end{itemize}
 
In both cases some external hardware is required, the amount of which will only fully 
understood after a more detailed evaluation. A possible outcome could be that a small 
non-volatile programmable device could provide the most flexible solution, but the effects 
of radiation need further examination.

Larger devices with non-volatile configuration memories could be used for the entire 
front-end logic, but apart from the increased cost and limited re-programming cycles making 
these less desirable, these devices are not able to re-program themselves. 

Reliability in the detector environment of these memories over the lifetime of an experiment, 
as well as SEUs, also need to be addressed. Non-volatile memories have limited programming 
cycles, whereas modern SRAM-based FPGAs boast that they can re-configured an infinite number 
of times. By refreshing the configuration periodically, the effect of SEU corruption can be 
minimised. Considering an FPGA with one million logic gates (which specifies a 4 Mbit PROM), configuring this device 
at 40\,MHz would take 100\,ms - a re-config every 250\,ms bunch-train!

Being able to re-configure the FE when needed also allows the selection of a smaller 
component that can be configured for specific function, rather than a large one-size fits 
all device.

Reducing the number of components on the front-end-module is advantageous from a cost, 
reliability and dead material point of view, so we will focus on methods of generating a 
data-stream onto the detector that requires the minimum of components to extract signals from 
the physical network interface module.

{\bf Clock and Control}
To ensure the on-detector electronics captures data from actual collisions, all components 
in the detector are synchronised to the bunch-crossing clock. Bunch-train start, stop and 
ID signals are also be required. Traditionally this is done with a bespoke, low-latency, 
highly synchronous system, but here we will look at using a commercial network switch.

Switches are not designed for synchronous signal fanout. In fact the now obsolete forerunner 
to the switch, the hub, is much better suited to the task. Modes for signal broadcasting do 
exist, but these need to be closely examined. Studies on latency and port-to-port skew will 
be undertaken.

To obtain maximal control over timing, the lowest level protocols will need to be used. 
Directly accessing the network at the physical layer (i.e. as a simple serial link) would 
facilitate complete control over data-packet composition and timing. For this a specialist 
and customisable network interface card is required.

To regenerate the clock at the detector end, the time structure of the data transmitted needs 
to be structured accordingly. This would probably take the form of 
a local oscillator being periodically resynced to the data clock. As the bunch-crossing 
interval is 7\,ns, a 1\,Gbit (1\,ns) link would be the minimum rate needed. Board-level signals, 
such as train-start may also need to be decoded directly from the data stream (as with the 
failsafe system above).

In summary, this is as an investigation to verify that a clock, control 
and configuration system can be constructed using commercial network hardware for 
distribution.

\subsection{Off-detector receiver}

An ideal system would have all data from the detector for each bunch train sent directly 
to a single PC where full event reconstruction would be performed. However, considering 
that the electromagnetic calorimeter alone will contain 24 million channels, even after data 
suppression this seems unfeasible.

It would, however, be desirable to geographically group as many channels as possible at 
a processor. This would permit broader local clustering which could be used as an input 
to full event reconstruction later. The fundamental questions to be answered are how much 
of the calorimeter can be received into one PC and how much needs to be received to make 
local clustering effective. The impact this has on full event reconstruction and on 
simulated physics needs to be evaluated. This task therefore involves a combination of 
hardware and simulation work. 

\begin{figure}[htp]
\begin{center}
~\epsfig{file=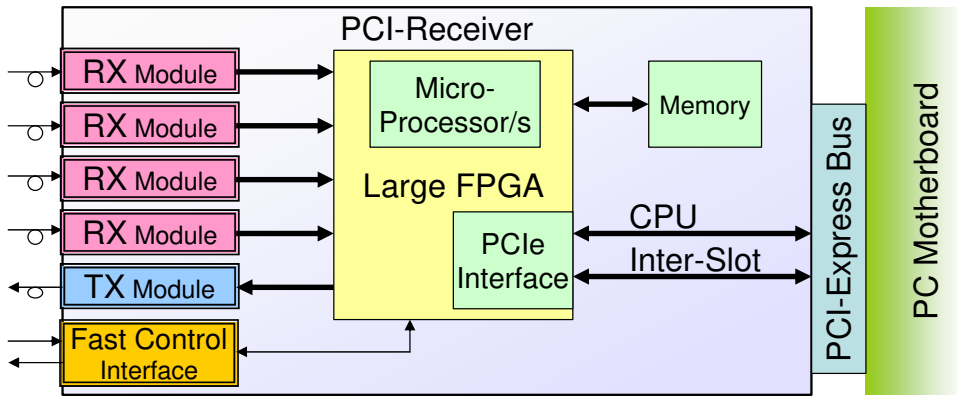,height=6cm}
\caption{Design of a customised PCI card.}
\label{fig:pci_card}
\end{center}
\end{figure}

We propose a system comprising PCI cards (see Fig.~\ref{fig:pci_card}) mounted in PCs. 
These cards will provide timing, clock, control and configuration data, be capable of performing 
local clustering, as well as  being able to be re-configured to act as data transmitters 
to exercise the receivers. We will use the ``new'' PCI Express bus standard ~\cite{PCIexpress} 
as this offers increased bandwidth and flexibility by using many high bandwidth serial links 
(called lanes) instead of a single parallel bus. It is expected to be scalable due to 
future increases in the number of lanes. Aside from providing an interface to the processor, 
lanes can be setup between cards, allowing advanced clustering prior to processing.

Although it is hard to predict what technology will be available for the final system, 
gaining experience with serial bus technologies now, will prove valuable for estimates 
of how much data can be realistically processed by a computer and help us understand the 
benefits and bottlenecks, as well as the scalability.

Much of this DAQ workpackage relies on novel methods in using network infrastructure, so a 
 card providing both direct access to the network hardware and room to build any 
other functionality is useful. This applies to the PCI Express interface too, where the 
card will be able to test-drive the bus and provide debug feedback and monitoring.

\section{A model DAQ system for the ECAL}
\label{sec:model}

As has been discussed in this note many options exist for the 
functionality of the components at various stages. It is therefore 
difficult to make a definitive statement on what a DAQ system would like 
and how many components it would have. However, work is ongoing to try and 
do as much of the work (data reduction) as possible on the detector, 
thereby reading out as little as possible. Therefore we consider that the 
ADC and thresholding are done in a 32-channel VFE and buffering in the FE.   
Under this model and making reasonable assumptions on the data rate a 
model system is detailed below.

The raw data size is assumed to be 2\,Bytes per channel with an additional 
4\,Bytes per channel needed for a timing label. Imposing the threshold 
suppression reduces the data rate by a factor of 100, leading to a data 
size for the whole detector of ($24 \times 10^6 \times 4886 \times 6)/100 =$ 
7.0\,GBytes. The size per slab is then 1.1\,MBytes. As the data is buffered 
in the FE, it is read out between bunch trains giving a rate per slab of 4.7\,MBytes/s.   
Early testing will readout some slabs individually, but as the system  
grows and stabilises, it would clearly be advantageous to bundle the data 
from several slabs together. Bundling half a tower of 20 slabs would give a 
rate of 0.75\,Gbit/s which could be adequately read out using current fibres. 
Such a grouping would lead to 300 fibres coming off the detector. Assuming a 
PCI card which can receive 8 fibres, this would require 38 PCI cards. With a
PC hosting 2 such cards than a system of 19 PCs would be required. We then  
assume a 20\% contingency for more fibres reading out neighbouring areas to allow 
for redundancy and for faulty PCs or PCI cards. This gives a total of about 
50 PCI cards housed in 25 PCs a clearly manageable number for such a granular 
calorimeter.

\section{Summary}
\label{sec:summary}

A conceptual design of a data acquisition system for the ILC calorimeter has been 
discussed. The concept relies heavily on commercial equipment and is generic such that it 
could be applied to other detector systems. Potential bottlenecks have been identified and 
form a programme of research and development of the next three to four years for the 
CALICE-UK groups. Bench tests and real-time tests with prototype calorimeters,  
within the EUDET project, will be undertaken. After this period, a technical design of the 
data acquisition system will be possible.



\end{document}